\documentclass[prb,twocolumn,floatfix,showpacs,preprintnumbers, 
               superscriptaddress,amsmath,amssymb]{revtex4}
\usepackage[dvips]{graphics}
\usepackage{psfrag}
\usepackage{graphicx}
\usepackage{epsf}
\usepackage{ucs}
\usepackage[utf8]{inputenc}

\begin{document}
\title{Shot noise in transport through ``double quantum dots"}

\author {Jasmin Aghassi} 
\affiliation{Forschungszentrum Karlsruhe, Institut f\"ur Nanotechnologie,
76021 Karlsruhe, Germany}
\affiliation{Institut f\"ur Theoretische Festk\"orperphysik,
Universit\"at Karlsruhe, 76128 Karlsruhe, Germany}
\affiliation{Center for Functional Nanostructures (CFN), Universit\"at
  Karlsruhe,  76128 Karlsruhe,Germany}
\author {Axel Thielmann} 
\affiliation{Forschungszentrum Karlsruhe, Institut f\"ur Nanotechnologie,
76021 Karlsruhe, Germany}
\affiliation{Institut f\"ur Theoretische Festk\"orperphysik,
Universit\"at Karlsruhe, 76128 Karlsruhe, Germany}
\affiliation{Center for Functional Nanostructures (CFN), Universit\"at
  Karlsruhe,  76128 Karlsruhe,Germany}
\author {Matthias H. Hettler} 
\affiliation{Forschungszentrum Karlsruhe, Institut f\"ur Nanotechnologie,
76021 Karlsruhe, Germany}
\author {Gerd Sch\"on} 
\affiliation{Forschungszentrum Karlsruhe, Institut f\"ur Nanotechnologie,
76021 Karlsruhe, Germany}
\affiliation{Institut f\"ur Theoretische Festk\"orperphysik,
Universit\"at Karlsruhe, 76128 Karlsruhe, Germany}
\affiliation{Center for Functional Nanostructures (CFN), Universit\"at
 Karlsruhe,  76128 Karlsruhe,Germany}

\date{\today}

%%%%%%%%%%%%%%%%%%%%%%%%%%%%%%%%%%%%%%%%%%%%%%%%%%%%%%%%%%%%%%%%%%%%%%%%%%%%%%
\begin{abstract}
Motivated by activities of several experimental groups we investigate electron transport through two coherent, strongly coupled 
quantum dots (``double quantum dots"), 
taking into account both intra- and inter-dot Coulomb interactions.
The shot noise in this system is very sensitive 
to the internal electronic level structure of the coupled dot system and its 
specific coupling to the electrodes. 
Accordingly a comparison between experiments and our predictions should allow
for a characterization of the relevant parameters.
We discuss in detail the effect of asymmetries, either asymmetries 
in the couplings to the electrodes or a detuning of the quantum dot levels out
of resonance with each other. 
In the Coulomb blockade region super-Poissonian noise appears 
even for symmetric systems. For bias voltages above the sequential 
tunneling threshold super-Poissonian noise and regions of negative 
differential conductance develop if the symmetry is broken sufficiently strongly.  
% either 
%for asymmetric dot-electrode couplings or for non-resonant dot levels with 
%symmetric couplings.
%Our predictions should allow qualitative determination of 
%the internal energy level structure of coupled quantum dots as well as 
%their coupling strength to the leads. 
\end{abstract}

\maketitle
%%%%%%%%%%%%%%%%%%%%%%%%%%%%%%%%%%%%%%%%%%%%%%%%%%%%%%%%%%%%%%%%%%%%%%%%%%%%%%

\section{Introduction} 
While early studies of electron transport through mesoscopic systems such 
as quantum dots or molecular systems concentrated on the current~\cite{sohn-etal,wiel}, 
more recent activities, both experimental~\cite{safonov,nauen,fujisawa,onac}
and theoretical~
\cite{bulka,elattari,thielmann,haug-kiesslich,belzig1,jordan,oppen1,flindt,dong,thielmann_co,chinese,Aghassi_3level}, include the analysis of shot noise. The 
latter provides additional insight into the quantum transport 
properties~\cite{blanter} and allows a more detailed characterization of
%, e.g., quantum information and nanomechanical systems, as well as molecular 
the quantum transport device.

For `local' systems, such as single (multilevel) quantum dots, above the sequential 
tunneling threshold the shot noise power $S$ is typically sub-Poissonian, implying that 
the Fano factor $F=S/2eI $, where  $I$ is the mean
current, is less than unity.  If the level couplings are asymmetric, e.g. in
the presence of magnetically polarized electrodes,
the noise can become super-Poissonian. In this case the Fano factor takes values larger
than unity~\cite{bulka,thielmann}. Very recently it was found that 
enhanced noise can also be found in symmetric systems inside the Coulomb 
blockade region where the current is much 
suppressed~\cite{belzig1,belzig2,thielmann_co}.
On the other hand, for 'non-local' systems, such as serially coupled 
quantum dots, due to their
complex internal level structure, super-Poissonian noise can develop even in
fully symmetric situations and above the sequential 
tunneling threshold~\cite{Aghassi_3level}. 
These 'non-local' systems exhibit a pronounced and sensitive dependence of
their transport characteristics on internal parameters and couplings. 

In this article we study sequential transport in a system of two strongly coupled
quantum dots. Specifically, we consider a double quantum dot (DQD) in series
in which the left dot is coupled weakly to left electrode and similarly
the right dot to the right electrode, while the two
dots are coupled strongly via electron tunneling, and they also interact 
electrostatically via the Coulomb interaction 
 (see Section~\ref{sec:model}). 
Our  results (in Section~\ref{sec:results})  address two distinct issues:

(i) In Section~\ref{subsec:symm} we study the shot noise of the
symmetrically coupled DQD in the Coulomb blockade regime, 
generalizing the work of Ref.~\onlinecite{belzig2}. Co-tunneling processes are assumed to be weak,  
hence transport is due only to thermally activated processes.
We find that in the Coulomb blockade regime the relation between two energy
scales, the sequential tunneling energy $\epsilon_{seq}$ and the difference of
the first excitation energy and the ground state (which is also the inelastic
cotunneling energy) $\epsilon_{co}$, determines the occurrence or absence of
super-Poissonian noise. This part of our analysis is valid generally for 
weakly coupled system in the Coulomb blockade regime, for any value of the gate voltage, 
as it  depends only on generic properties of the 
internal electronic structure of the interacting dot system.

(ii) Above the sequential tunneling threshold, for the weakly coupled
DQD  super-Poissonian noise can only appear if the left $\leftrightarrow$ right 
symmetry is broken. For a non-local
system like the DQD this symmetry breaking can be achieved in two
qualitative different ways: (a) The symmetry of the electrode-dot
couplings is broken, while the DQD is unchanged,
see Section \ref{subsec:asymm_coup}. Here, the step positions
in the current and noise characteristics are not influenced by the asymmetry. 
But for sufficient asymmetry in the coupling  negative
differential conductance (NDC) appears, i.e. the current decreases with
increasing bias, while at even stronger asymmetry additionally super-Poissonian
noise appears. (b) The symmetry of the electrode-dot
couplings is preserved, while the symmetry of the DQD-Hamiltonian is
broken by detuning the dot level energies, see 
Section \ref{subsec:asymm_level}. Here, the step positions
in the current and noise characteristics differ for different 
degrees  of detuning. The DQD eigenfunctions
become spatially non-uniform which breaks the parity symmetry of the
effective coupling of the various eigenstates to the electrodes. This, in turn,
leads again to NDC and eventually with
further detuning to super-Poissonian noise.

As such asymmetries are easily detected 
in an experiment, we can learn more about the underlying asymmetries 
of the couplings, electronic structure and also the total spin 
of the states participating in transport. 
In the available measurements setups on DQDs of various 
groups~\cite{kotthaus,marcus,tarucha,ensslin} 
metallic finger gates allow for controlled manipulation
of the relevant parameters, e.g  the  
electrostatic potential of the 
individual dots as well as the inter-dot and dot-electrode couplings.
%~\cite{kotthaus,marcus,tarucha,ensslin}.
While it might be difficult to unambiguously distinguish the two cases of 
symmetry breaking, our predictions should still be verifiable 
in shot noise measurements 
in available systems. We also  comment and compare our
results to related theoretical work~\cite{elattari,dong,chinese} on
similar systems in  { Section~\ref{subsec:comparison}}. 
We conclude with a summary in   { Section~\ref{sec:summary}}.

\section{Model and technique}
\label{sec:model}
We consider two coupled quantum dots, each  with a sufficiently large
level spacing such that we can restrict ourselves to one spin-degenerate
level per dot. Including electron hopping between the dots as well as intra-dot and
inter-dot (nearest neighbor) Coulomb interactions we arrive at the    
Hamiltonian 
$\hat H = \hat H_{\rm L} + \hat H_{\rm R} + \hat H_{\rm DQD} + 
 \hat H_{\rm T,L} + \hat H_{\rm T,R}$
with
\begin{eqnarray}
&& \hspace*{-0.8cm} \hat H_{r} = \sum_{k \sigma}\epsilon_{k \sigma r} a_{k
  \sigma r}^{\dag} a_{k \sigma r},\,\,
\hat H_{{\rm T},r}= \sum_{i k \sigma}(t_r a_{k \sigma
            r}^{\dag} c_{i \sigma} 
            + h.c.), \nonumber \\
&& \hspace*{-0.8cm}\hat H_{\rm DQD} = \sum_{i \sigma} \epsilon_{i} n_{i \sigma}  
  - t\sum_{\sigma} 
( c_{1 \sigma}^{\dag}c_{2 \sigma} +h.c. ) \nonumber \\  
&& \hspace*{+0.2cm} + \,U\sum_i n_{i\uparrow}n_{i\downarrow}  +U_{nn}\sum_{\sigma \sigma'} n_{1\sigma}n_{2\sigma'},
\label{hamilton}
\end{eqnarray}
where $i=1,2$ denote the dot levels and $r={\rm L},{\rm R}$. 
Here, $\hat H_{\rm L}$ and $\hat H_{\rm R}$ model the non-interacting
electrons with density of states 
$\rho_e= \sum_k \delta(\omega -\epsilon_{k \sigma r})$ 
in the left and right electrode ($a_{k \sigma r}^{\dag}, a_{k \sigma r}$ are 
the Fermi operators for the states in the electrodes). The chemical potentials 
(Fermi energy) of the electrodes $\mu_{\rm L},\mu_{\rm R}$ 
in equilibrium set the zero point of the energy scale. The term
$\hat H_{\rm DQD}$ describes the coupled dots with on-site energy 
$\epsilon_i$ and  inter-dot hopping 
$t$. $c_{i\sigma}^{\dag}, c_{i\sigma}$ are Fermi operators for the molecular 
levels, and $n_{i\sigma}=c_{i \sigma}^{\dag}c_{i \sigma}$ is the number 
operator. The strength of the intra-dot and  inter-dot Coulomb repulsion is 
given by $U$  and $U_{nn}$ respectively. The parameters $U$  and
$U_{nn}$ can be related to the charging energies of the dots and the
various capacitances when comparing to experimental setups as
described e.g. in Ref.~\onlinecite{wiel}.
Other electron interaction terms could be considered by much more
elaborate models, as done in Ref.~\onlinecite{hettler_prl} 
for computation of the $I-V$ characteristics.
For the  effects on the shot noise that we wish to study, 
the simpler  model above suffices.
The terms $\hat H_{{\rm T},{\rm L}}$ and $\hat H_{{\rm T},{\rm R}}$
describes tunneling between the leads and the corresponding adjacent
dot. The respective coupling strength is characterized by the 
intrinsic line width 
$\Gamma_r = 2 \pi |t_r|^2 \rho_e$, where $t_r$ are the
tunneling matrix elements.

We are interested in transport through the DQD, 
in particular in the current $I$ and the (zero-frequency) current noise $S$.
They are related to the current operator 
$\hat I = (\hat I_{R} - \hat I_{\rm L})/2$, with 
$\hat{I}_r = -i(e/\hbar) \sum_{i k \sigma} \left( t^r_i a_{k \sigma r}^{\dag} c_{i\sigma} -  h.c.\right)$
being the current operator for electrons tunneling into lead $r$, by 
$I = \langle \hat{I} \rangle$ and 
\begin{equation}
S = \int_{-\infty}^{\infty} dt \langle \delta \hat{I}(t) \delta \hat I(0) 
+ \delta \hat{I}(0) \delta \hat I(t) \rangle
\label{noisedef}
\end{equation}
where 
$\delta \hat I(t)=\hat I(t)-\langle \hat I \rangle$.

We compute transport via a master equation for the diagonal elements of the
reduced density matrix of the DQD system. This approach
has been discussed in detail in Ref.~\onlinecite{thielmann}. For
completeness, here we summarize the most salient aspects of this approach.
The reduced density matrix is expressed in the eigenstate basis  of the
dot Hamiltonian $H_{DQD}$, Eq.\ref{hamilton}. For the $N=2$ dot system
there exist $4^N=16$ eigenstates $\chi$ of the form $\chi=\sum_{s}c_s
|s\rangle$, where $|s \rangle$ denotes a basis state of the form
$|n_{1\uparrow}n_{1\downarrow}n_{2\uparrow}n_{2\downarrow}\rangle$ and the
$c_s$ are the corresponding coefficients. The
analytic form of the eigenstates and eigenvalues of our Hamiltonian can be
found in Ref.~\onlinecite{bulka2}. 
We calculate the transition rates ${\bf W}$ between the eigenstates via 
perturbation theory (in this case 
via the Fermi golden rule) in the electrode-DQD couplings  $\Gamma^r$.
The bold face indicates matrix notation related to the eigenstate
labels $\chi$.
In the stationary situation (no explicit time  dependence of the bias) 
the density matrix becomes time independent and  we can find
the average occupation of the eigenstates, i.e. stationary
probabilities ${\bf p^{st}}$  by the solution of the  master rate equation
\begin{equation}
\sum_{\chi'}W_{\chi, \chi'}p^{st}_{\chi'}=0.
\label{eq:master}
\end{equation}
under the condition that $ \sum_{\chi} p^{st}_{\chi}= 1$.
For the calculation of the current $I$ and current noise $S$, we use the
diagrammatic technique on the Keldysh contour developed in 
Ref.~\onlinecite{diagrams} which was expanded for the description 
of the noise in Ref.~\onlinecite{thielmann}.
In first order perturbation theory, the current and shot noise are given by:
\begin{equation}
  I = {e\over 2\hbar} {\bf e}^T {\bf W}^{I}
  {\bf p}^{{\rm st}}
\label{eq:I1}
\end{equation}
\begin{equation}
  S = {e^2\over \hbar} {\bf e}^T \left( 
    {\bf W}^{II} {\bf p}^{{\rm st}} + {\bf W}^{I} 
    {\bf P} {\bf W}^{I} {\bf p}^{{\rm st}} \right).
\label{eq:S1}
\end{equation}
The vector $\bf e$ is given by $e_\chi = 1$ for 
all $\chi$. The  objects ${\bf W}^{I} ({\bf W}^{II})$ denote the Laplace
transform of the transition rates (in the time domain) 
between eigenstates $\chi$ with one (two)
current vertex(ices) due to $\hat{I}_r$ replacing a tunneling vertex 
due to $H_{\rm T,L}$ or 
$H_{\rm T,R}$. The "propagator" $ {\bf P} $ can be found from the
Dyson equation. In first order perturbation theory it is obtained from the
equation
\begin{equation}
{\bf P}={\bf \tilde W}^{-1}{\bf Q,}
\label{eq:prop}
\end{equation}
where ${\bf \tilde W}$ is identical to the transition rates matrix $\bf W$
defined above but with one row (arbitrarily chosen) $\chi_0$ being replaced
with $(\Gamma, ....,\Gamma)$ 
and
$Q_{\chi\chi'}=(p^{st}_{\chi'}-\delta{\chi'\chi})(1-\delta_{\chi'\chi_0})$,
see Ref.~\onlinecite{thielmann} for the full details.

We point out that expression Eq. \ref{eq:S1}
consists of two terms.  The first term,  denoted by 
$S_{irr}$ is due to  ${\bf W}^{II}$, i.e.
the  noise diagrams with two current vertices in a single irreducible block. 
%This term will at least give a noise of size $e I$. 
The second term,  denoted by $S_{red}$,  is due to reducible noise diagrams, 
i.e. diagrams with a "propagator"  ${\bf P}$ between the two  current 
vertices of Eq.~\ref{noisedef}  at different times. 
It therefore accounts for the 
electronic structure and the correlations of the system.
It is mostly the second term $S_{red}$ that is responsible for the 
interesting correlation physics and super-Poissonian noise, 
as we will see below.

\section{Results}
\label{sec:results}
In the following we discuss current and shot noise for systems described by a 
Hamiltonian of the type of Eq.~(\ref{hamilton}) in first order perturbation theory in the
tunnel couplings $\Gamma_r$. In the first part of the discussion special 
emphasis is put on examining the behavior
of the Fano factor (Noise) in the Coulomb blockade region. The second part 
will be devoted to the discussion of asymmetry effects induced to
the double dot system by asymmetric coupling to the leads or detuned level 
energies, respectively. In the case of symmetric couplings we choose 
$\Gamma_{\rm L} = \Gamma_{\rm R}=2.5 {\rm \mu eV} $ defining a total line width of   
$\Gamma= \Gamma_{\rm L} + \Gamma_{\rm R} = 5 {\rm \mu eV}$. 
(We choose this explicit energy scale as we are varying a number of 
different energy parameters in the following.)
Our perturbation expansion is valid for temperatures much larger 
than the tunnel couplings. 
Throughout this paper, we choose $k_{\rm B} T = 10 \Gamma$ which translates to
$ T=50 {\rm \mu eV} \sim 0.6  {\rm  K}$.
The dot system  is characterized by the level energies $\epsilon_i$, 
the intra-dot 'Hubbard' repulsion $U$, and the nearest neighbor charge
repulsion $U_{nn}$. If not stated otherwise the level energies are chosen to be resonant, 
$\epsilon_1=\epsilon_2=\epsilon$.

A current is driven by an applied bias voltage $V_{\rm b}=\mu_{\rm L}-\mu_{\rm R}$. 
We assume the voltages to drop symmetrically and, since the dot-electrode coupling is weak compared to the dot-dot coupling, entirely at the  electrode-dot tunnel junctions. 
This implies that the level energies of the dots are independent of 
the applied voltage. 
%(This situation differs from the one realized in the experiments of van der Wiel et al.~\cite{wiel} were the interdot tunnel coupling was weak.)
% even if the couplings $\Gamma^r_l$ are not symmetric. 
Effects such as level detuning  due to asymmetric or incomplete 
voltage drops and or applied gate voltages 
could easily be included. We do not consider these effects 
here, as they add unnecessary complexity to the results presented below. 
We include only a single level per dot (plus interactions), assuming that the 
level spacing within each dot is larger 
than all other energy scales. 

To proceed we diagonalize the dot Hamiltonian  $H_{DQD}$
including the interaction terms. The resulting 
eigenstates can be organized according to the two quantum numbers: total
charge $-qe$ (with $q$ an integer, $q \in 0,1,2,3,4$) and total spin 
(singlets, doublets and triplets for our DQD model)\cite{bulka2,note2}.
As the onsite energies $\epsilon_i$ are decreased to lower, negative values
(experimentally achieved by a gate voltage applied to both dots)
the ground state charge shifts from  $q=0$ to increasing values $q=1$, $q=2\ldots$.
% at values when the lowest states of neighboring charge sectors are degenerate. 
While previous work~\cite{elattari,dong} has focused mostly on the zero 
charge ($q=0$) ground state we study the more interesting case (see
below)  with  a ``half filled'' ground state ($q=2$), where the low-bias 
transport sensitively depends on the 
spatial and spin structure of the eigenstates in the various charge sectors.
   
For the sequential transport in quantum dot systems at low bias 
two energy scales are relevant: 
(1) The "sequential energy gap" $\epsilon_{seq}$ denotes the energy difference
between the ground state with charge $-qe$ and the first excited states 
with the charge  $-(q+1)e$ (`anion ground state') {\it or} with 
charge $-(q-1)e$ (`kation ground state'), depending on which one is smaller.
The sequential tunneling threshold, i.e. the bias above which the current is
no longer suppressed due to Coulomb blockade, is reached at
$V=2 \epsilon_{seq}/e$ for symmetric bias.
(2) The energy gap between the ground state with
charge $-qe$ and the first excited state with the same charge,
denoted in the following by $\epsilon_{co}$. The energy $\epsilon_{co}$
is also known as the `vertical' gap, and is related to the HOMO-LUMO gap in
molecular systems. It would be the energy scale relevant for inelastic 
co-tunneling processes. Note, however, that co-tunneling processes, 
which are of second order in $\Gamma_r$, are not included in this work. 
If one would start with a ground state of zero charge ($q=0$) the energy scale
$\epsilon_{co}$ would not exist within our model, due to the restriction to single level dots. 
As we consider the case of a half filled ground
state we avoid such an artefact. 
Also note that recent experiments on double quantum dot systems 
with applications for quantum computing~\cite{marcus,tarucha,Petta}
work with ground states of non-zero charge.

\subsection{Symmetrically coupled quantum dots} 
\label{subsec:symm}
\begin{figure}[h]
%\centerline{\includegraphics[width=8.5cm,angle=0]{graphics/half_fil1.eps}}
\centerline{\includegraphics[width=8.5cm,angle=0]{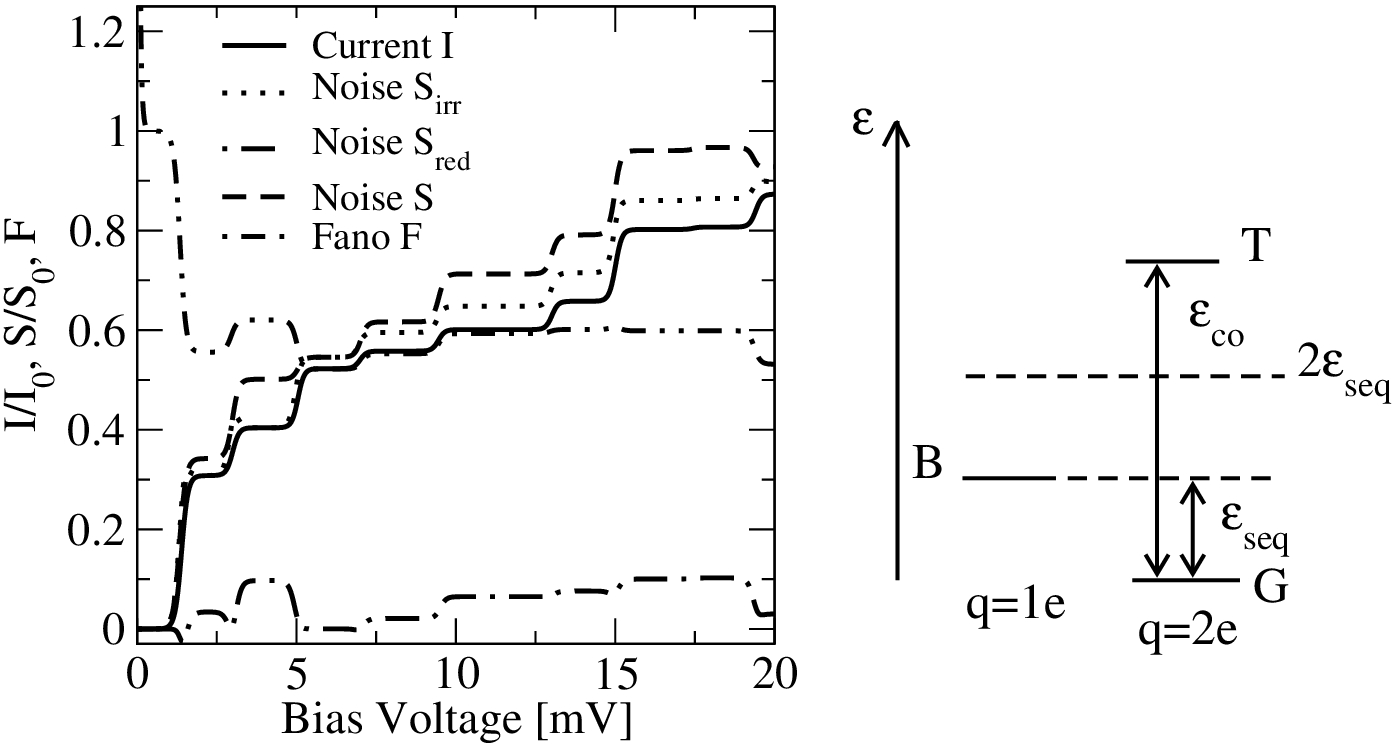}}
\caption{ {\it Left panel}: Current $I$ and shot noise $S$ vs. bias voltage 
for  a double dot system 
with $k_{\rm B} T=0.05$,  $t=2$, $U=10 $, $U_{nn}=5$ and 
$\epsilon=-5.5$ resulting in a doubly occupied ground state ($q=2e$), all units
in meV. 
The noise $S$ is sub-Poissonian for 
all bias voltages. This is always the case if the first vertical excitation
energy is larger than twice the sequential tunneling threshold, 
$\epsilon_{co} > 2 \epsilon_{seq}$, see the sketch in the right panel.
 Current and noise curves are normalized to 
$I_0=(e/\hbar)2\Gamma$ and 
$S_0=(e^2/\hbar)2\Gamma$, respectively. {\it Right panel}: 
sketch  of the low  energy spectrum. The nature of the states G,T and
B is discussed in the text.}
\label{fig:half_fil1}
\end{figure}

We begin with a system of two dots in series and energy parameters 
such that the DQD is half filled in the ground state (no bias applied).
In the right panel of Fig.\ref{fig:half_fil1} part of the energy excitation
spectrum resulting from the diagonalization of the Hamiltonian is displayed. 
The ground state G is a singlet state (total spin $0$) and charge 
$-2e$ ($q=2$). It is delocalized over the two dots
(a combination of the  four two electron singlet basis vectors $| s \rangle$) 
with an eigenvalue $E_{\rm G}$ dependent on all parameters
of $H_{DQD}$,  $E_{\rm G}= 2 \epsilon +1/2( U + U_{nn} - \Delta)$, 
where~\cite{bulka2} $\Delta=\sqrt{16 t^2 + (U-  U_{nn})^2}$. 
The first excited state is the bonding state B with $q=1$. 
It is a doublet with total spin ${1}/{2}$, eigenvalue
$\epsilon-t$, and is also  delocalized over both dots. Therefore,
the energy scale $\epsilon_{seq}$ is given 
$\epsilon_{seq}=E_{\rm B}- E_{\rm G}$.
The second excited state is a triplet (total spin $1$) with $q=2$. 
In the triplet state one electron each is "fixed" to one 
dot. Therefore, its eigenvalue is independent of the inter-dot hopping $t$ and 
the on-site repulsion $U$. 
the energy scale $\epsilon_{co}$ is therefore given by 
$\epsilon_{co}=E_{\rm T}- E_{\rm G}$. 
The rest of the spectrum is not shown, since for the following discussion 
we will refer to a
bias regime for which other states are not yet important. The higher excited
states are responsible for the step features above $V_{\rm b} \sim 5 mV$.
Note that in the artificial limit $ U \rightarrow \infty$ the energy
scale $\epsilon_{co}$ vanishes. In this
case, the triplet and singlet states would be degenerate and 
some of the effects described below would disappear.
%The resulting plots for current, shot noise and Fano factor $F=S/2 e I$
%are displayed in the following figures
%(Figs. \ref{fig:half_fil1} - \ref{fig:half_fil3}).

Fig.\ref{fig:half_fil1} shows the typical behavior for a fully symmetric system
with $\epsilon_{co}>2 \epsilon_{seq}$ : both current and noise rise 
monotonically in steps, while the Fano factor will fall
between values of 1 (Poissonian noise) and 1/2 (symmetric double barrier noise)
for the large bias region, i.e. a bias voltage larger than all 
excitation energies. In general, the Fano factor will 
not fall with a monotonous dependence on the bias. 
This non-monotonicity is due to
the second term in the noise expression Eq. \ref{eq:S1}, 
%(dotted line in Fig. \ref{fig:2D-filled}), 
associated with the propagator {\bf{P}}, which can give positive 
and negative contributions (it is negative in the entire 
Coulomb blockade region and becomes positive only on the first plateau). 
In the Coulomb blockade current and noise are both (equally) exponentially
suppressed resulting in a Fano factor of Poissonian value. 
At small bias, $eV_{\rm b} \ll k_{B}T$, 
the noise is dominated by thermal noise,
described by the well known hyperbolic cotangent behavior which leads to a
divergence of the Fano factor \cite{blanter,loss}.

\begin{figure}[h]
%\centerline{\includegraphics[width=8.5cm,angle=0]{graphics/half_fil2.eps}}
\centerline{\includegraphics[width=8.5cm,angle=0]{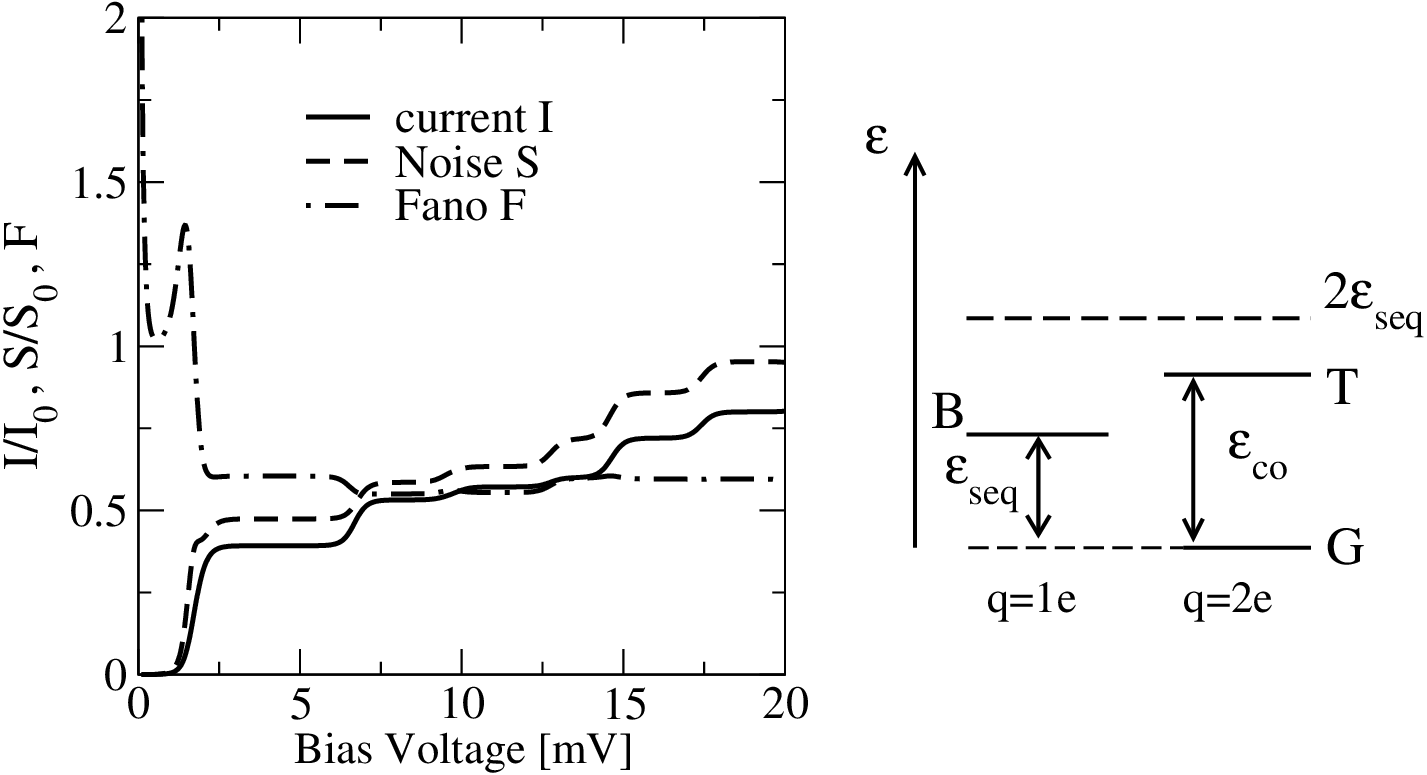}}
\caption{ {\it Left panel}: Current $I$ and shot noise $S$ vs. voltage for a 
double dot system 
with $k_{\rm B} T=0.05$,  $t=2$, $U=12$, $U_{nn}=4 {\rm meV}$ and 
$\epsilon=-5.3$.Super-Poissonian noise (Fano factor $F > 1$) develops in
the Coulomb blockade regime. {\it Right panel}: low energy spectrum, where now
$\epsilon_{seq}<\epsilon_{co}<2\epsilon_{seq}.$}
\label{fig:half_fil2}
\end{figure}

\begin{figure}[h]
%\centerline{\includegraphics[width=5.5cm,angle=0]{graphics/log_half_fil2.eps}}
\centerline{\includegraphics[width=5.5cm,angle=0]{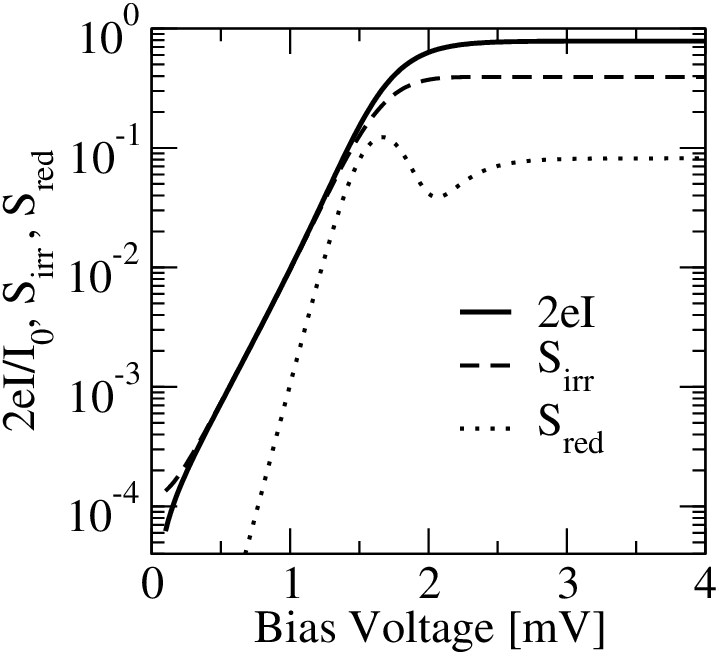}}
\caption{Enlarged low bias region of Fig.\ref{fig:half_fil2}. $2eI, S_{red}$ and
$S_{irr}$ are plotted semi-logarithmically. The first part
of the noise $S_{irr}$ grows with bias as the current, providing a Poissonian
noise contribution. The second part of the noise $S_{red}$ becomes appreciable
for $V  > 2(\epsilon_{co}-\epsilon_{seq})/e$ and causes the total noise enhancement.}
\label{fig:log_half_fil2}
\end{figure}

\begin{figure}[h]
%\centerline{\includegraphics[width=6.50cm,angle=0]{graphics/half_fil3.eps}}
\centerline{\includegraphics[width=6.50cm,angle=0]{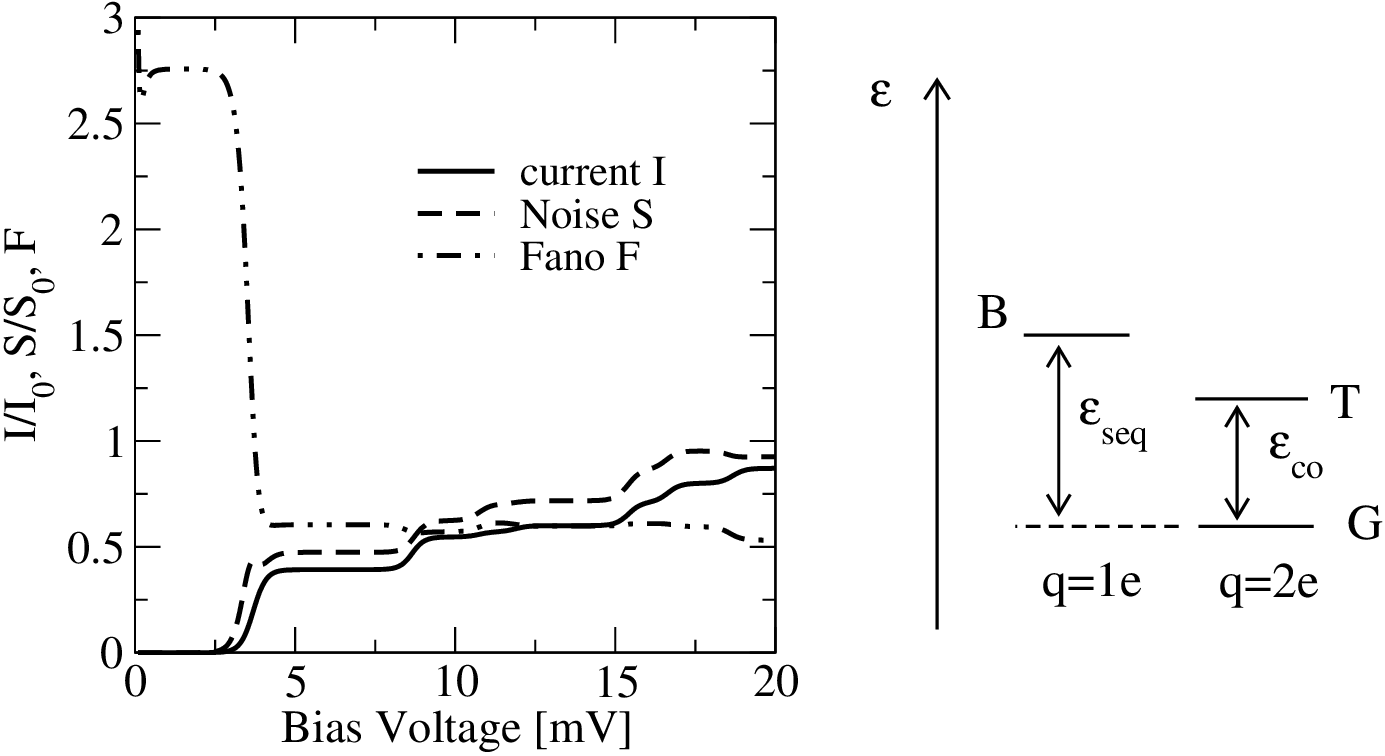}}
\caption{ {\it Left panel}: Current $I$ and shot noise $S$ vs. voltage for a 
double dot system with $k_{\rm B} T=0.05$,  $t=2$,  
$U=12$, $U_{nn}=4 $ and $\epsilon=-6.3$. Super-Poissonian noise
develops in  the entire Coulomb blockade
region.  {\it Right panel}: the corresponding low energy spectrum, 
where $\epsilon_{co}<\epsilon_{seq}$.}
\label{fig:half_fil3}
\end{figure}

If we now lower the onsite energy $\epsilon$ we energetically favor states 
with larger charge and thus increase the energy $\epsilon_{seq}$ as 
compared to 
the situation as shown in the right panel of Fig.\ref{fig:half_fil1}, 
while preserving the energy $\epsilon_{co}$.
Thereby, we can realize a situation in which 
$\epsilon_{seq}<\epsilon_{co}<2\epsilon_{seq}$, see Fig.\ref{fig:half_fil2}. Above the sequential threshold
the current and noise curve look very similar to the situation in 
Fig. \ref {fig:half_fil1}, with the expected small shifts in the step
positions. However, in the
Coulomb blockade region the Fano factor behaves differently to before. 
After the region of thermal
noise accompanied with divergent Fano factor, a Poissonian value of $F=1$ is
reached. For higher bias and close to (but still below) the 
sequential tunneling threshold a
peak like feature (actually a short plateau) appears in the Fano factor.
This is caused by a relative enhancement of the noise, visible in
Figs. \ref{fig:half_fil2} and \ref{fig:half_fil3} 
by the apparent shift of the noise curve to
lower bias in the left panel. 
The increase of the Fano factor is due to the second term in the noise 
expression $S_{red}$ (Eq. \ref{eq:S1}), see the
Fig.\ref{fig:log_half_fil2} (note the semi-logarithmic scale). 
The first part of the noise $S_{irr}$  
provides the finite thermal noise around zero bias. It then 
grows with bias with the same exponential behavior as the current and 
contributes a Poissonian term $2eI$ to the shot noise. 
In contrast, the (now positive) second 
part $S_{red}$ becomes only appreciable for a bias 
$V_{\rm b} >(\epsilon_{co} -\epsilon_{seq})/e$ and renders the shot noise
super-Poissonian above this bias. This noise enhancement is due to
the possible thermal occupation and subsequent sequential depletion of 
excited states that lead to small cascades of tunneling events interrupted 
by long (Coulomb) blockages. The alternation of these processes with different
time scales results in a noisy current. Consequently, the
Fano factor is larger than unity, indicating super-Poissonian noise. 
This effect was recently discussed in some detail by Belzig and
co-workers \cite{belzig1,belzig2}, for systems restricted to  
a singly occupied ground state. At a bias higher than the sequential threshold
the noise recovers sub-Poissonian behavior.

For the same parameters as above but with further lowered onsite energy
$\epsilon=-6.3$ we obtain a situation where $\epsilon_{co}<\epsilon_{seq}$. 
The current, noise and Fano factor for such a situation are depicted in
Fig. \ref{fig:half_fil3}. For a bias larger than the sequential tunneling
threshold the curves show again generic behavior as displayed in 
Fig. \ref{fig:half_fil1} and Fig.\ref{fig:half_fil2}. 
However, in the Coulomb blockade regime
and after divergent thermal noise behavior we directly obtain a 
super-Poissonian Fano factor $F\approx 2.8$ in form of a plateau  
and do not recover a Poissonian value in the entire Coulomb blockade regime
at all. In this case, $S_{red}$ gives a large contribution that behaves with 
the same exponential behavior as the current 
rather than dropping faster than the current at low bias, as in 
Fig.~\ref{fig:half_fil2}). Thus the noise  is enhanced in the entire
Coulomb blockade regime. The term $S_{irr}$ again provides the thermal noise
at very low bias and a contribution of $2eI$ below the bias  
$V_{\rm b} >(\epsilon_{seq} -\epsilon_{co})/e$. Above this bias, there is 
a redistribution between $S_{irr}$ (losing) and $S_{red}$ (winning), however,
the sum of the two terms grows exactly like the current, leading to a constant
(super-Poissonian) Fano factor.

Summarizing the above discussion we can distinguish three 
possible situations in the Coulomb blockade region: \\
i) For $\epsilon_{co} >  2 \epsilon_{seq}$
the sequential processes start at a bias before the excited states come 
into play, and the noise is Poissonian, i.e. $F=1$ once the thermal noise 
becomes negligible.
This is the case for Fig. \ref{fig:half_fil1}, as sequential transport via the
ground state G and the ``bonding state'' B  takes place before 
the triplet state T  can be reached from the bonding state B.\\
ii) For  $ \epsilon_{seq} <\epsilon_{co} <  2 \epsilon_{seq}$ there is
super-Poissonian noise $F>1$ in the bias range 
$2(\epsilon_{co}-\epsilon_{seq})/e < V_{\rm b} < 2\epsilon_{seq}/e$, 
see Fig. \ref{fig:half_fil2}. This is due to 
the transport scenario discussed above, as for a bias in this range a 
thermally excited system can  for a time  do sequential transport through 
the excited states, before recovering to the ground state.\\
iii) For $\epsilon_{co} < \epsilon_{seq}$ we have $F>1$ for the entire Coulomb
blockade region. For a bias $2(\epsilon_{seq}-\epsilon_{co})/e < V_{\rm b} <
2\epsilon_{seq}/e$ the situation is the same as in scenario ii). Below 
this bias range the physical picture due to
Ref.\onlinecite{belzig2} needs to be modified, 
as sequential transport is "blocked" (thermally activated) even out of the
first excited state for  $V_{\rm b} < 2(\epsilon_{seq}-\epsilon_{co})/e$. 
Nevertheless, the Fano factor actually remains constant as the bias drops
below  $ 2(\epsilon_{seq}-\epsilon_{co})/e$, see Fig. \ref{fig:half_fil3}.

%At this bias the relative size of the first and second noise terms of 
%Eq. \ref{eq:S1} changes, i.e the first term increases, 
%whereas the second term decreases accordingly. Both terms are exponentially
%suppressed in the same way as the current, different from the situation ii)
%where the reducible noise contribution vanishes faster than the irreducible
%term and the current, see the right panel of Fig. \ref{fig:half_fil2}.

However, as was pointed out in Ref.~\onlinecite{thielmann_co}, 
the super-Poissonian noise behavior 
due to sequential tunneling processes in the
Coulomb blockade regime is easily modified by co-tunneling processes. 
As elastic co-tunneling provides a Poissonian process are often much larger
current (in the Coulomb blockade region) than the exponentially small
sequential current, adding all contributions gives a Fano factor of 
nearly unity deep in the blockade region. 
Inelastic co-tunneling processes, on the other hand, can provide a ``true'' 
super-Poissonian noise around a bias $\epsilon_{co}/e$ 
above which the inelastic processes 
compete with the Poissonian elastic co-tunneling processes.
%corresponding to the first inelastic co-tunneling energy.
Therefore, if co-tunneling processes dominate sequential processes 
(typically for ratios above $\Gamma/T \sim 10^{-3}$ \cite{thielmann_co}) 
there is either no super-Poissonian noise, 
if $\epsilon_{co} >  2 \epsilon_{seq}$ as in scenario i), or, 
there is super-Poissonian noise starting
around a bias of $\epsilon_{co}/e$, similar to scenario ii).
The experimental distinction of scenarios ii) and iii) can therefore be 
difficult: although the Fano factor looks different in pure sequential
transport, if co-tunneling processes play a role, 
scenarios ii) and iii) will display qualitatively similar Fano factor behavior.

\subsection{Asymmetric dot-electrode couplings}
\label{subsec:asymm_coup}

We now turn to the discussion of transport above the sequential 
tunneling threshold, i.e. in the bias region where electrons can tunnel
sequentially through the DQD because they have sufficient energy to overcome 
the Coulomb blockade. For the symmetric situations as discussed above, the 
current and the noise increase monotonically in steps, where the step
positions are determined by the many-body excitations of the DQD.
For our DQD system, the noise in a symmetric transport situation remains
sub-Poissonian (Fano factor $F< 1$) at all bias above the sequential 
tunneling threshold.

\begin{figure}[h]
%\centerline{\includegraphics[width=8.5cm,angle=0]{graphics/asymmetric.eps}}
\centerline{\includegraphics[width=8.5cm,angle=0]{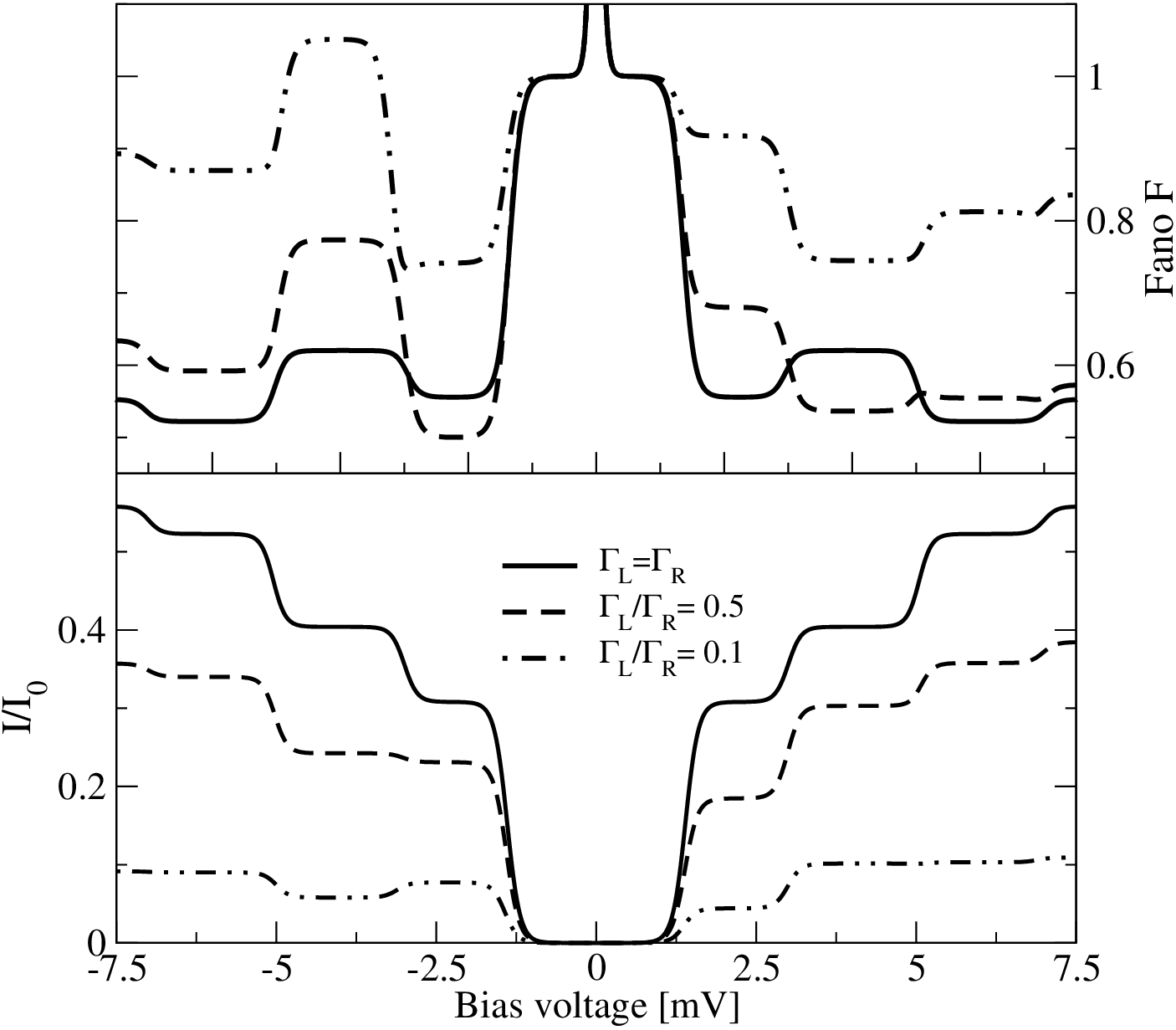}}
\caption{Current $I$ (absolute value) and Fano factor $S$ vs. voltage
  for asymmetric coupling
  in a double dot system with $k_{\rm B} T=0.05$,  $t=2$,  
$U=10 $ and $U_{nn}=5 $, $\epsilon=-5.5$. For strong asymmetry 
negative differential conductance and super-Poissonian noise appear 
only for negative bias voltages. Note that due to the asymmetry  the total
  line width  $\Gamma= \Gamma_{\rm L} + \Gamma_{\rm R}$ and the 
current are reduced relative to the  symmetric case.
}
\label{fig:asymmetric}
\end{figure}

This is changed in  situations with asymmetric couplings,
e.g. when the coupling to the left electrode is
suppressed relative to the coupling to the right electrode, 
${\Gamma_{\rm L}}/{\Gamma_{\rm R}}<1$.
As all energy parameters are chosen to be the same as in the
situation displayed in Fig.~\ref{fig:half_fil1} the ground 
state is again a two electron state with
$\epsilon_{co}>2\epsilon_{seq}$. Hence for the following discussion one
should refer to the qualitative energy spectrum shown in the right panel 
of Fig.~\ref{fig:half_fil1}. In Fig.\ref{fig:asymmetric} the
upper graph depicts the Fano factor and the lower graph the
absolute value of the current for various asymmetry ratios 
${\Gamma_{\rm L}}/{\Gamma_{\rm R}}$ 
(the current is negative for negative bias).
%given in its absolute value for negative and positive voltage. 
In the symmetric case, represented by the solid line, 
the Fano factor as well as absolute current and the
noise (that is not depicted here) show a fully symmetric behavior under the 
reverse of the bias voltage. The first plateau is reached when the 
transition from the doubly occupied ground state G ($q=2e$) to the lowest 
single occupied state, the bonding state B, ($q=1e$) becomes allowed 
at the sequential tunneling threshold ($V_{\rm b}=2\epsilon_{seq}/e)$).
At these plateaus the current, noise and Fano factor are functions of the
coupling constants $\Gamma_r$ only. At negative bias on the first plateau,
the Fano factor is given by 
\begin{equation}
F=\frac{4\Gamma_{\rm L}^2+
\Gamma_{\rm R}^2}{{(2\Gamma_{\rm L}+\Gamma_{\rm R})}^2}\, . 
\label{eq:fano}
\end{equation}
This gives a value of $\frac{5}{9}$ at the first plateau for 
symmetric coupling. For positive bias voltage one needs to 
exchange $\Gamma_{\rm L}$ with
$\Gamma_{\rm R}$, respectively. This result can be related to previous work 
by some of us\cite{note_fano}.  
%where the various plateau values of a 
%single dot system
%have been connected with the excitation energies of the
%dot spectrum and whether or not they lie within the energy window 
%defined by the applied bias. 
%In our case, the first plateau is due
%to transitions from the half filled ground state G ($q=2$) to the 
%singly occupied bonding state B ($q=1$). 

For the curves with ${\Gamma_{\rm L}}/{\Gamma_{\rm R}} \neq 1$ there is a 
clear  asymmetry in current and Fano factor. The first plateau value of the
Fano factor is increased for positive bias and (for smaller asymmetry) 
decreased for negative bias according to the above expression for the Fano 
factor. Further suppression of the left coupling leads to a region of 
negative differential
conductance (NDC) and eventually a super-Poissonian Fano factor on the second
plateau at negative bias (see dash-dotted curve for 
$\Gamma_{\rm L}/\Gamma_{\rm R}=0.1$). 
The reason for the current suppression and asymmetric behavior is the 
interplay of the asymmetric couplings and the internal electronic structure. 
The occupation of the states participating in transport at the plateaus is
highly sensitive to the asymmetric couplings.

Let us consider the first plateaus (positive and negative bias) of 
the current in the case $\Gamma_{\rm L}/\Gamma_{\rm R}=0.1$. 
For negative bias, in contrast to the symmetric case where the ground
state G and the bonding state B are equally occupied, we
have now a higher probability to be in the state G than in the
state B. This is due to the fact that it is ``easy''
to populate the DQD from the right but ``difficult''  to depopulate the 
DQD in direction of the left electrode because of the suppressed
coupling. As a consequence the system is occupied  by two electrons 
most of the  time. The reverse holds for positive bias, where the dot is 
most often occupied by one electron and consequently the 
probability to be in the state 
B on the first plateau is higher than to be in the ground state G. 

To obtain the  current $I$ we need to consider the
relevant current rates $W^I$ in addition to the probabilities of the various 
states. On the first plateaus, the relevant current rates 
$W^I_{{\rm G}\rightarrow  {\rm B}_{\sigma} }$ 
from ground state to bonding state(s) (with given spin $\sigma$) 
at negative bias are equal in magnitude to the reverse rates 
$W^I_{ {\rm B}_{\sigma}\rightarrow {\rm G} }$  
at positive bias (independent of the spin $\sigma$ of ${\rm B}_{\sigma}$).
Solving the master equation, as a result of the coupling asymmetry 
the probability $p_G$ to be in state G on the first plateau at negative bias is
however larger (almost twice) than  the occupation $p_{{\rm B}_{\sigma}}$ for 
states ${\rm B}_{\sigma}$  on the first plateau at positive bias. 
%However, accounting for the relevant spin multiplicities 
%(G is a singlet, B is a doublet), 
The combination of the same relevant current rate
% ($ | W^I_{{\rm G}\rightarrow {\rm B}} | = 
%| W^I_{{\rm B} \rightarrow {\rm G}  }|$) 
but different occupations %$p_{\rm G} > p_{{\rm B}_{\sigma}}$ 
leads to a higher (absolute) value of the current on the first plateau at 
negative bias than on the corresponding plateau at positive bias. 
To be concrete, if we consider the currents on the left interface of the DQD 
we have at negative bias a current with absolute value 
$| 2 \, W^I_{{\rm G}\rightarrow {\rm B}}\, p_{\rm G} |$ which is 
almost twice as large (for $\Gamma_{\rm L}/\Gamma_{\rm R}=0.1$) 
than the current 
$ W^I_{{\rm B}\rightarrow {\rm G}}\, 2 \,p_{{\rm B}_{\sigma}}$ 
going through the left interface at positive bias. Here, 
the factors of two originate from the spin summation over the
bonding state doublet.

On the first plateau the Fano factor is monotonically
increasing for positive bias with decreasing $\Gamma_{\rm L}/\Gamma_{\rm R}$
until it would reach the Poissonian value $F\rightarrow 1$ 
for very large asymmetry, resembling the noise of an effective single barrier.
For negative bias on the first plateau, the Fano factor 
(given by Eq. \ref{eq:fano}) is not monotonic: it
first decreases until it reaches $F=1/2$ for 
$\Gamma_{\rm L}/\Gamma_{\rm R}=0.5$, then it increases until it also would 
reach the Poissonian value for large asymmetry. This non-monotonic behavior 
reflects the interplay of asymmetry and 
different spin multiplicity of the relevant states G and B.

At the second plateau the transition from the bonding state B to the first
excited triplet state T ($q=2$) becomes possible and thus provides a second
current channel. The stationary probabilities are redistributed in the 
following way: For negative bias, the states G and each of the three 
triplet states ${\rm T}_m, m \in -1,0,1$ have approximately equal 
occupation (within 10 percent). 
As a consequence of the threefold spin multiplicity 
of the triplet the probability of the ground state 
decreases to less than one third of its value on the first plateau.
The bonding state B also loses some of its (already small) 
probability to the competing triplet states. The tunneling processes
from the triplet state(s) T to the 
bonding state(s) B contribute an additional current via the current rates
 $W^I_{{\rm T}\rightarrow {\rm B}}$ (per triplet state and spin of B).
However, even when summing over all the triplet contributions the resulting 
current is too small to compensate the loss from the processes involving the 
ground state. Therefore
a region of negative differential conductance (NDC) appears as soon 
as the triplet states play a role in transport for negative bias 
(at the considered coupling asymmetry). The NDC is accompanied 
by a (relatively) enhanced noise because the  competing processes involving the
ground state  and the triplets have sufficiently unequal rates  
$W^I_{{\rm G}\rightarrow  {\rm B}}$ and $W^I_{{\rm T}\rightarrow {\rm B}}$ 
to form 'slow' and 'fast' transport channels. This competition leads to the 
super-Poissonian Fano factor as  depicted in Fig. \ref{fig:asymmetric}. 

\begin{figure}[b]
%\centerline{\includegraphics[width=8.5cm,angle=0]{graphics/var_epsilon.eps}}
\centerline{\includegraphics[width=8.5cm,angle=0]{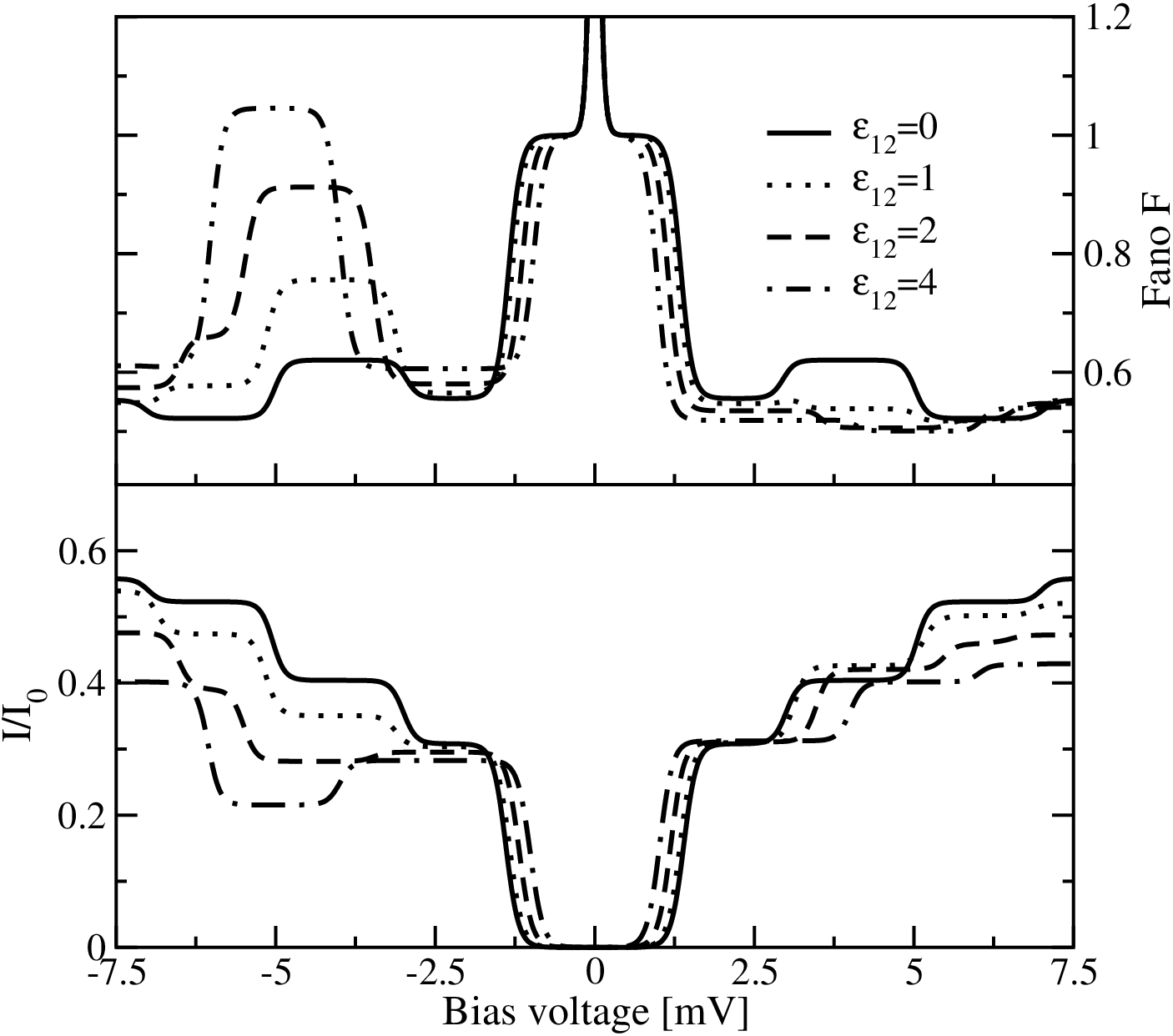}}
\caption{Current $I$  (absolute value) and Fano factor $S$ vs. voltage for 
various values of
  level detuning $\epsilon_{12}$ but with symmetric couplings to the leads
  and other energy parameters equal to the situation depicted in 
Fig.\ref{fig:half_fil1}.
 Stronger detuning $\epsilon_{12}$ leads to 
NDC and eventually super-Poissonian noise. In contrast 
to Fig.\ref{fig:asymmetric}, the bias (energy) positions of 
current and  noise  features are changed due to the modified dot Hamiltonian.}
% $k_{\rm B} T=0.005 {\rm meV}$,  $t=-0.2$,  
%$U=1.0 {\rm meV}$ and $U_{nn}=0.5 {\rm meV}$, $\epsilon=-0.55 {\rm meV}.$
\label{fig:var_epsilon}
\end{figure}

For positive bias on the second plateau the situation is quite different. 
Here,
%the bonding state B remains highly occupied, whereas the triplets 
%and the ground state have  small probabilities.
%( a factor $\sim 100$ smaller). 
the DQD remains mostly in the bonding state B, i.e. there is no 
major loss  of occupation for the bonding state (about 10 percent).  
Again the current leaving the dot  consists of two additive contributions.
In addition to the ground state contribution already present 
on the first plateau, the transitions between bonding to the triplet 
states add a large contribution 
and thus the current increases stepwise to a second, higher plateau. 

The above illustrates that although for one bias direction (here
positive) the current and shot noise show generic behavior 
(and the Fano factor is always sub-Poissonian), 
the situation can be quite different for the reverse 
bias. As such asymmetries are easily verified in an experiment, we
can learn much about the underlying asymmetries of the couplings and 
the spin multiplicities of the states participating in transport.
Note that the NDC and super-Poissonian noise would completely disappear if
we would take the onsite Coulomb repulsion $U \rightarrow \infty$. 
%The different behavior of the singlet ground state and the triplet 
%states results from the fact that 
Due to a finite $U$ the singlet ground state can benefit from  
'local singlets', i.e. states with two electrons of opposite spin on 
the same dot, whereas there is no equivalent for triplet states.  
Therefore, the singlet ground state
has a lower energy and different transitions rates as 
compared to the triplet states, both of which are necessary conditions for the
NDC and super-Poissonian noise in the considered single-level model.

\subsection{Detuned level energies}
\label{subsec:asymm_level}
The discussion above serves as a basis to qualitatively understand transport 
in the more complicated situation when the symmetry of the DQD Hamiltonian
itself is broken, rather than merely its coupling to the electrodes.
In the following, we detune the level energies
$\epsilon_1-\epsilon_2=\epsilon_{12}$ and also vary the inter-dot hopping 
$t$ while the other parameters of the dot system remain the same and the
couplings remain symmetric, $\Gamma_{\rm L}=\Gamma_{\rm R}$.
For an experiment, this implies a gate electrode for each dot that can be 
controlled separately. Similar to above  in Fig.~\ref{fig:asymmetric}, 
if roughly  $|\epsilon_{1} -\epsilon_{2}|=|\epsilon_{12}| > |t|$, 
NDC and super-Poissonian noise can be realized at some bias.

In Fig. \ref{fig:var_epsilon} we show current and Fano factor for different 
level detuning $\epsilon_{12}$.
The black solid line corresponds to symmetric
couplings and resonant levels $\epsilon_1=\epsilon_2$. It is the same as
depicted in Fig.\ref{fig:asymmetric}. If we start detuning the levels, i.e.
$\epsilon_{12}\ne0$ we change our excitation
energies and the states become more localized on the dot with lower energy
(here the right dot). Consequently, we find the 
current and Fano factor plateaus at different (energy) positions as before,
with different length of the plateaus. 
Note that the current on the first plateau only weakly changed for 
all $\epsilon_{12}$ considered here. 
%Furthermore, it behaves almost symmetrically under the reverse of bias. 
This is due to the fact that despite of the changed
Hamiltonian, the tunneling rates from ground state to bonding state 
as well as the occupations of these states are almost the
same. The occupations on the first plateau are also only 
weakly dependent on the sign of the bias, quite different to the situation
with asymmetric couplings considered above.
Only with an even stronger level detuning would the current
be significantly changed on the first plateau. 
However, the considered detuning of levels still leads to NDC and 
eventually to a super-Poissonian Fano factor, e.g. for $\epsilon_{12}=4)$ at 
negative bias. The effect of triplet states on the second plateau
is qualitatively the same as in 
the scenario with asymmetric coupling discussed above.  
For positive bias the current remains monotonic and the Fano factor 
sub-Poissonian. In agreement with previous results \cite{elattari}
the maximum current at very large bias (not shown)
decreases with increased detuning although the total coupling $\Gamma$ 
remains unchanged.

\begin{figure}[h]
%\centerline{\includegraphics[width=8.5cm,angle=0]{graphics/var_t_E02.eps}}
\centerline{\includegraphics[width=8.5cm,angle=0]{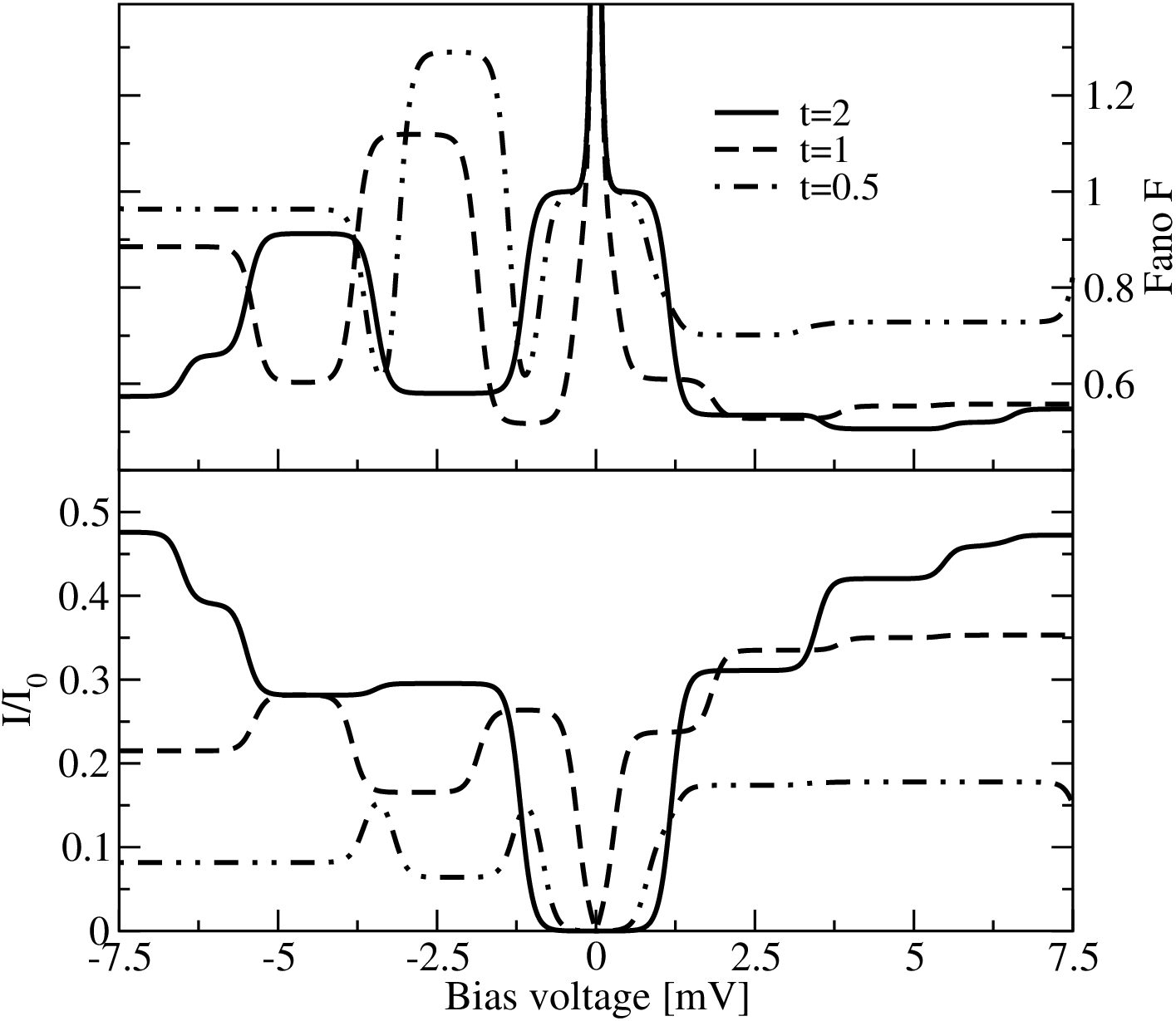}}
\caption{Current $I$  (absolute value) and Fano factor $S$ vs. voltage for 
various ``hopping''   parameters $t$ and a level detuning 
of $\epsilon_{12}=0.2$, symmetric coupling to the leads and otherwise 
same parameters as in the situation 
depicted in Fig. \ref{fig:half_fil1}. Reduced hopping causes a smaller 
total current although super-Poissonian
noise and NDC develop similarly as in Fig.\ref{fig:var_epsilon}.}
% $k_{\rm B} T=0.005 {\rm meV}$,  $t=-0.2$,  
%$U=1.2 {\rm meV}$ and $U_{nn}=0.4 {\rm meV}$, $\epsilon=-0.55 {\rm meV}.$
\label{fig:var_t1}
\end{figure}

Instead of further increasing the level detuning one can also achieve 
"localization" of states by  decreasing the inter-dot 
hopping $t$. Let us consider again the symmetrically coupled system 
($\Gamma_{\rm L}=\Gamma_{\rm R}$) at a fixed detuning of
$\epsilon_{12}=2$ for various values of the inter-dot hopping $t$ (see
Fig.\ref{fig:var_t1}). The solid line corresponds again to the case, in which
$\epsilon_{12}=2$ and $t=2$, 
as was also depicted in Fig.\ref{fig:var_epsilon}(dashed line).
As expected, the plateaus of the current are again asymmetric since we have
detuned level energies. If we now decrease $t$, the bonding state  and the 
ground state will be separated by only a very small energy (as $U_{nn}=5$ and 
$(\epsilon_{1}+\epsilon_{2})/2=5.5$)
and thus the Coulomb blockade almost disappears.
For positive bias both current and Fano factor (noise)  behave generically.
%i.e. increase in a stepwise manner and the latter is always sub-Poissonian.
The first plateau for negative bias is again due to tunneling processes
involving the states B and G. 
At the second plateau the triplet T  starts participating
in the transport and is strongly occupied, resulting in NDC and
super-Poissonian noise as discussed
above. At even more negative bias there exists a second 
region of NDC (for the  cases $t=1$ and $t=0.5$). 
This is where the anti-bonding state (not
depicted in the spectrum in the right panel of 
Fig. \ref{fig:half_fil1}) is also 
contributing to the transport. The maximum current at large bias (not shown) 
depends on the inter-dot hopping $t$~\cite{elattari} if the dot 
levels are out of resonance.

From Fig. \ref{fig:var_epsilon} and Fig. \ref{fig:var_t1} one can conclude
that a higher degree of localization of the states participating in transport,
achieved either by a strong detuning of level energies or a decrease in the 
inter-dot hopping, favors transport features such as NDC 
and  makes the current more and more noisy, leading eventually to 
super-Poissonian noise. 
Reducing the hopping (at fixed detuning) therefore has a 
similar effect on transport  as a larger detuning at fixed hopping.
However, as the DQD spectrum differs non-linearly between different 
parameter sets with identical ratio $\epsilon_{12}/t$ the resulting transport
curves can not be scaled, but depend explicitly on the value of 
each parameter. 

\subsection{Comparison with related theoretical work}
\label{subsec:comparison}

\begin{figure}[h]
\vspace{0.5cm}
%\centerline{\includegraphics[width=8.5cm,angle=0]{graphics/var_t_E0.eps}}
\centerline{\includegraphics[width=8.5cm,angle=0]{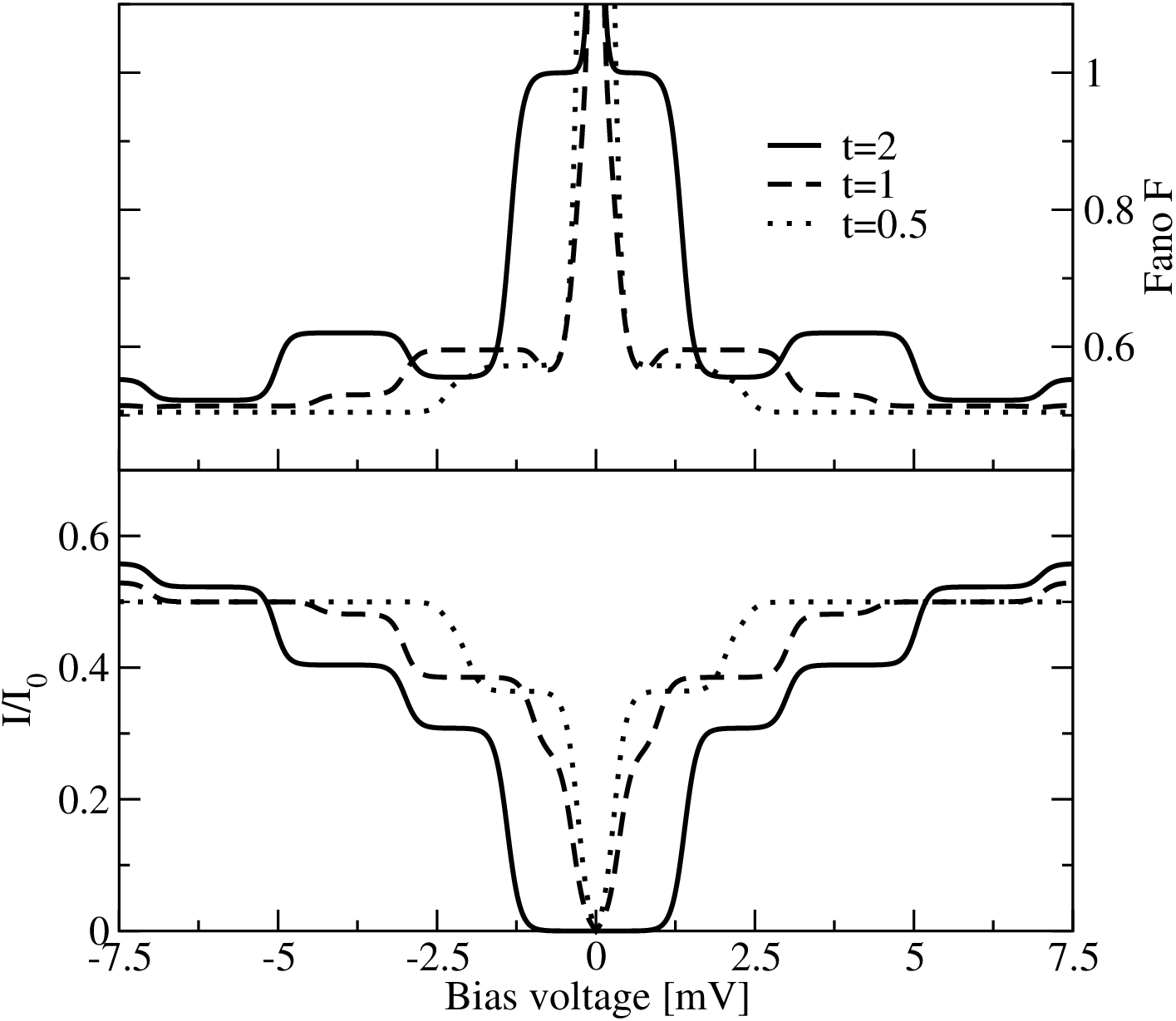}}
\caption{Current $I$  (absolute value) and Fano factor $S$ vs. voltage 
for different values of
``hopping'' $t$ without level detuning ($\epsilon_{12}=0$) for symmetric 
couplings and same parameters as in the situation depicted in Fig. 
\ref{fig:half_fil1}. Note that current 
and Fano factor are both symmetric under the reverse of bias voltage,
since there is no source of asymmetry.
% $k_{\rm B} T=0.005 {\rm meV}$,  $t=-0.2$,  
%$U=1.2 {\rm meV}$ and $U_{nn}=0.4 {\rm meV}$, $\epsilon=-0.55 {\rm meV}.$
}
\label{fig:var_t2}
\end{figure}

For reference we show in Fig.~\ref{fig:var_t2} the current and Fano factor
for a fully symmetric system, i.e. equal couplings to left and right and
resonant level energies but for different values of the inter-dot hopping
$t$. As expected all curves behave symmetric under the
reverse of bias. Similar to  Fig. \ref{fig:var_t1}, for smaller hopping 
$t$ the sequential tunneling threshold, determined by the energy 
distance of the states G and B, becomes very small for the chosen parameters 
and thus the Coulomb blockade almost disappears. 
Since there is no asymmetry in the
system, not in the couplings, nor in the energy levels, we do not expect and
do not find regions of NDC and/or super-Poissonian noise. This is specific
to this DQD system, in which there are only interfacial
dots and thus there is always a finite probability for the
electrons to depopulate the dot structure. In contrast, 
in chains of three and more quantum dots we
do find super-Poissonian noise even for a fully symmetric system 
\cite{Aghassi_3level}, if the ratio $U_{nn}/{t}$ is sufficiently large. 
In these more complex systems the enhancement of 
the noise is due to a combination of the non-local many-body wave functions
and the states with higher total spin ($3/2$).

Note that although the current (Fano factor) does very much depend 
on the value of the hopping $t$ for the low bias regime depicted 
in Fig.~\ref{fig:var_t2}, the maximum current at very large bias (not shown)
is actually independent of the hopping $t$. This is due to our neglect of
off-diagonal matrix elements of the reduced density matrix.
Such  off-diagonal elements have been included in the work of
Ref.~\onlinecite{elattari} by Ellatari and Gurvitz. 
They   
%We conclude with a comparison of our work with the results of earlier works 
%\cite{elattari,dong,chinese}. 
%In Ref. \onlinecite{elattari} Elattari and Gurvitz
have studied the  shot noise through two
coupled quantum dots via a quantum rate equation, i.e. a master
equation involving also the off-diagonal elements of the 
reduced density matrix. 
The effect of off-diagonal elements on transport are negligible for 
the weak coupling situation we consider, but become increasingly important, if
the coupling $\Gamma$ becomes comparable to the intrinsic energy scales of the
dot system, such as the hopping energy $t$. Naturally, a straight 
perturbative approach as ours does not make
sense for such large $\Gamma$ (at least not to first order in $\Gamma$), so
our results apply to the explicit case of weak coupling, such that $\Gamma$ is
the smallest energy scale. The approach of Ref.~\onlinecite{elattari},
introduced in detail in Ref.~\onlinecite{gurvitz}, is not explicitly
restricted to a small $\Gamma$. 
%because it deals non-perturbatively (in a
%sense) with the coupling to the leads. 
Under certain assumptions and restrictions the contributions from all 
the lead states can be 
"integrated out" and a quantum rate equation (still linear in  $\Gamma$) 
for the relevant parts of the reduced density matrix is obtained.  
After solving the quantum rate equation, the current and the shot noise can be
computed.

However, it is important to note that the approach of  
Ref.~\onlinecite{elattari,gurvitz} 
imposes severe restrictions on the transport situations it can treat. 
The applied bias has to be 
very large, such that all states (or excitations) of the dots that are 
considered for transport lie well in 
between the Fermi levels (chemical potentials) of the electrodes. 
On the other hand, the states 
not considered for transport should be very far away from the chemical
potentials of the leads. Figuratively speaking, the approach of 
 Ref.~\onlinecite{elattari} applies to the center of a very long 
(in principle, infinitely long) plateau of the I-V characteristics. 
Such plateaus are not realized for the situations we considered, 
with the exception of 
the last plateau, when the bias is much larger than {\it all} excitations
(or energies) of the dot system. 
In Ref.~\onlinecite{elattari} the large bias
regime for a spinless version of our Hamiltonian (implying also $U=\infty$)
 and the cases  $U_{nn}=0$ as well as $U_{nn}=\infty$
were considered. For  $U_{nn}=0$ we have a non-interacting system of spinless
fermions, whereas for  $U_{nn}=\infty$ only states with at most a single 
(spinless) electron on the dot  system are  relevant. 
%(the empty dots state, the bonding and the anti-bonding state). 
The results of  Ref.~\onlinecite{elattari}  for current and shot noise 
for $\Gamma/t << 1$ approach our results (not shown), as it should be.
%For example, for $U_{nn}=\infty$ the Fano factor in the case of 
%symmetric coupling and aligned levels ($\epsilon_1 = \epsilon_2$) 
%comes out to be $5/9$. \cite{thielmann}
If  $\Gamma/t$ is not negligibly small,
corrections to our results become noticeable that reduce the current and 
the shot noise, but also the Fano factor.  If  $\Gamma >> t$ 
the system turns into an effective single barrier system, so that the 
current is $\propto t^2/\Gamma$ and the noise becomes Poissonian, 
$F \rightarrow 1$. 

Recently, two groups\cite{dong,chinese} claimed to have extended the approach
of Ref.~\onlinecite{elattari} to the arbitrary bias regime. The results of 
Ref.~\onlinecite{chinese} for the current on single dot systems 
are in perfect agreement with earlier work by us \cite{thielmann}. This 
is expected, as there are no possible "coherence effects" in a single level 
system, if spin is conserved in tunneling. 
Whereas Ref.~\onlinecite{chinese} does not discuss interacting double dot 
systems (or shot noise) Ref.~\onlinecite{dong} also considers the spinless
version of our DQD Hamiltonian.
They discuss both current and shot noise through a 
double dot system with resonant levels $\epsilon$ and an infinitely large 
intra-dot repulsion $U$, starting from an empty dot system as a ground state. 
Therefore, they can only consider two possible regimes with finite 
current. i) For a bias $2 \epsilon< eV_{\rm b} <2(\epsilon +U_{nn}) $, they consider
to have at most one electron on the double dot system. In this case they 
recover the result of Ref.~\onlinecite{elattari} for the case $U_{nn} =
\infty$. ii) For bias larger than all excitation energies 
($eV_{\rm b} > 2(\epsilon +U_{nn}$)) they recover 
the {\it non-interacting} result  of  Ref.~\onlinecite{elattari}. 
In contrast, in our approach there
are two additional possible transport situation that arise from the fact 
that there are two eigenstates in a singly occupied double dot system, 
namely the bonding (B) and the anti-bonding (A) state that differ by the energy
$2t$. Therefore, in general there will be two steps corresponding to
excitations of the bonding state ($\epsilon-t$) and the anti-bonding state 
($\epsilon+t$)
out of the ground state ($\epsilon=0$) and another two steps corresponding
to the excitation energies  to the doubly occupied state at 
$\epsilon+U_{nn}-t$ and $\epsilon+U_{nn}+t$. 
Hence, the current and noise
characteristics for the double dot system with $U\rightarrow \infty$ should
show four steps, unless broadening effects due to temperature or the line
width $\Gamma$ are so large that they smear out the steps. 
In Ref.~\onlinecite{dong}, Fig. 6, there are only two steps which are 
broadened by temperature only (via the Fermi functions). Therefore, 
at best it corresponds to a situation for which $t < \Gamma <<  k_{\rm B} T$, 
i.e. a situation in which the system resembles more an effective
single barrier system at high temperature.

The "concatenation" of two results by  Ref.~\onlinecite{dong} 
that are derived under the assumption of 
effectively infinite bias in  Ref.~\onlinecite{elattari}
also leads to the peculiar effect that the current
exhibits NDC behavior for a ratio of  $\Gamma/t > 2$. While
it is possible that a fully spatially symmetric system can display NDC 
\cite{Aghassi_3level},
%we wonder whether that the considered system of only two dots
%and  symmetric coupling lacks the necessary complexity to do so. \cite{note}
%(without breaking of the spatial symmetry). \cite{note}
in our weak coupling theory ($\Gamma << t$) of the DQD system 
NDC can only occur  with broken symmetry 
such as detuned level energies, asymmetric couplings etc. as discussed in 
this paper. \cite{note}  The NDC effect in  Ref.~\onlinecite{dong} occurs 
in a parameter regime where our theory clearly does not apply.
It would be interesting to see
whether other approaches like the ones based on equation of motion methods
\cite{bulka2,wacker} can confirm or disprove the NDC effect displayed in  
Fig. 6 of Ref.~\onlinecite{dong}. Very recent work of
Ref.~\onlinecite{wunsch} includes level renormalization terms
left out by  Ref.~\onlinecite{elattari,dong} that modify the current
characteristics qualitatively in the regime $\Gamma >> t $. This shows
that the inclusion of off-diagonal elements of the density
matrix is a rather delicate matter.\\

\section{Summary}
\label{sec:summary}
In summary we have discussed transport through a double quantum dot (DQD) 
system with both intra- and inter-dot Coulomb interactions in the 
sequential transport
picture, as currently studied by several experimental groups. We found that
the behavior of the shot noise in the Coulomb blockade is directly related to
the underlying low energy spectrum of the DQD system
characterized by two intrinsic energy scales, the sequential tunneling energy
$\epsilon_{seq}$ and the first vertical excitation energy out of the ground
state, $\epsilon_{co}$.
For a symmetric system in the Coulomb blockade we distinguished between three scenarios:
i) For a first vertical excitation energy that is smaller than twice the
sequential tunneling energy $\epsilon_{co}>2\epsilon_{seq}$ the Fano factor (noise) is always
sub-Poissonian, i.e. $F<1$, as sequential processes start before excited
states come into play.
ii) If $\epsilon_{seq}<\epsilon_{co}<2\epsilon_{seq}$ thermally activated sequential transport
leads to super-Poissonian Fano factors in the bias range
$2(\epsilon_{co}-\epsilon_{seq})/e < V < 2\epsilon_{seq}/e$. 
iii) For the case 
%even higher excitation energies, 
$\epsilon_{co}<\epsilon_{seq}$ the
Fano factor remains super-Poissonian in the 
entire Coulomb blockade regime. Our findings are valid for arbitrary ground
state charges and also apply to larger systems with more than two coupled dots, 
as they depend only on the above mentioned generic energy scales of the
interacting dot system.

%By inducing asymmetries in the system either by asymmetric couplings to the 
%electrodes or by detuning the quantum dot level energies out of resonance with
%each other, we furthermore found that a sufficiently strong symmetry 
%breaking leads to
%negative differential conductance and eventually to super-Poissonian 
%noise in a range of bias voltage  above the sequential tunneling threshold. 
Additionally, we discussed the effect of asymmetries in the system
realized by either asymmetric couplings to the 
electrodes or by detuning the quantum dot levels out of resonance with
each other. In the case of asymmetric dot-electrode couplings we 
obtained an asymmetric current voltage characteristics as has been observed in
experiments before. For very strong asymmetry negative differential
conductance and eventually super-Poissonian noise with Fano factors $F>1$
develop. These features develop at the same energy positions,
i.e. at the same bias voltage for any asymmetry ratio
$\Gamma_L/\Gamma_R$ since the DQD spectrum
remains unchanged. In contrast detuning the dot levels out of resonance also
leads to NDC and super-Poissonian noise for sufficiently strong asymmetry, but now
at voltages that depend on the strength of the asymmetry
as the DQD spectrum is changed.
These features only appear for one bias direction, $V<0$ or
$V>0$, depending on which coupling $\Gamma_r$ ($r$=right,left) is suppressed
or which quantum dot has a lower level energy.
Furthermore, we found that at a fixed detuning $\epsilon_{12}$ the  current 
is reduced with decreasing inter-dot hopping $t$. The latter results in a
stronger localization of states on individual dots similar to the case of
strongly detuned quantum dots.
Therefore a weaker inter-dot hopping and a stronger detuning at fixed inter-dot
hopping cause similar transport characteristics. 

To conclude, we have shown that transport properties of double quantum dots, in particular the shot
noise, show a strong sensitivity on the internal electronic structure and the
coupling strengths to the electrodes. This sensitivity should allow for a detailed characterization of
these energy scales in a given experiment.

{\em Acknowledgments.}
We acknowledge helpful discussions with Maarten Wegewijs, 
Bernhard Wunsch, Jonas Pedersen and Andreas Wacker, and the
financial support by the DFG via the Center for 
Functional Nanostructures (CFN).


\begin{thebibliography}{99}
\bibitem{note2}The magnetic quantum number, 
i.e. the z-component of the total spin plays no role in our 
spin degenerate DQD system without magnetic fields and 
normal metal electrodes.
\bibitem{note_fano} Our first plateau value corresponds to the
second plateau of Ref. \onlinecite{thielmann} 
(which considered spin non-degenerate  levels), 
with the further requirement that 
$\Gamma_{\rm L}$ and $\Gamma_{\rm R}$ need to be exchanged as the 
plateau is reached by a transition from a higher charge state 
to a lower charge state.
\bibitem{note} NDC can be achieved in local multi-level systems,
provided that one level couples sufficiently weaker than the other(s).
However, the considered non-local two dot system can not be mapped to such a 
system without breaking the spatial symmetry.
\end{thebibliography}
\end{document}